\documentclass[english,american,aps,prb,twocolumn,superscriptaddress,showpacs,showkeys]{revtex4-1}
\usepackage{amsmath,amssymb}
\usepackage{amsthm}
\usepackage{graphicx}
\usepackage{subfigure}

\usepackage[latin9]{inputenc}

\usepackage{babel}

\begin{document}

\title{Statistically optimal analysis of state-discretized trajectory data
from multiple thermodynamic states}

\author{Hao Wu, Antonia S.J.S. Mey, Edina Rosta, Frank No\'{e} \\ Correspondance to: frank.noe@fu-berlin.de}

\begin{abstract}
We propose a discrete transition-based reweighting analysis method
(dTRAM) for analyzing configuration-space-discretized simulation trajectories
produced at different thermodynamic states (temperatures, Hamiltonians,
etc.) dTRAM provides maximum-likelihood estimates of stationary quantities
(probabilities, free energies, expectation values) at any thermodynamic
state. In contrast to the weighted histogram analysis method (WHAM),
dTRAM does not require data to be sampled from global equilibrium,
and can thus produce superior estimates for enhanced sampling data
such as parallel/simulated tempering, replica exchange, umbrella sampling,
or metadynamics. In addition, dTRAM provides optimal estimates of
Markov state models (MSMs) from the discretized state-space trajectories
at all thermodynamic states. Under suitable conditions, these MSMs
can be used to calculate kinetic quantities (e.g. rates, timescales).
In the limit of a single thermodynamic state, dTRAM estimates a maximum
likelihood reversible MSM, while in the limit of uncorrelated sampling
data, dTRAM is identical to WHAM. dTRAM is thus a generalization to
both estimators. 
\end{abstract}
\maketitle

\section{Introduction}

The dynamics of complex stochastic systems are often governed by rare
events - examples include protein folding, macromolecular association,
or phase transitions. These rare events lead to sampling problems
when trying to compute expectation values from computer simulations,
such as molecular dynamics (MD) or Markov chain Monte Carlo (MCMC).

One approach to alleviate such sampling problems is to increase the
rate at which the rare events occur by generating and combining simulations
at different thermodynamic states. For example, proteins can easily
be unfolded at high temperatures, and protein-ligand complexes can
dissociate at artificial Hamiltonians, where protein and ligand have
reduced interaction energies. Generalized ensemble methods, such as
replica exchange molecular dynamics\foreignlanguage{english}{\cite{Greyer1991}},
parallel tempering\foreignlanguage{english}{\cite{Hukushimi1996,Hansmann1997,Sugita1999}}
and simulated tempering\foreignlanguage{english}{\cite{Marinari1992}}
exploit this observation by coupling simulations at different temperatures
or Hamiltonians within an MCMC framework. Yet another example is umbrella
sampling\cite{Torrie_JCompPhys23_187} which uses a set of biased
Hamiltonians to ensure approximately uniform sampling along a set
of pre-defined slow coordinates.

All of the aforementioned enhanced sampling methods are constructed
such that for long simulation times, the equilibrium distribution
of each thermodynamic state will be sampled from. With that in mind,
reweighting estimators make use of all simulation data by reweighting
each probability density from the thermodynamic state sampled at to
the condition of interest \emph{via} the Boltzmann density. The most
frequently used reweighting estimators are the weighted histogram
analysis method (WHAM) \cite{FerrenbergSwendsen_PRL89_WHAM,Kumar1992},
bin-less WHAM \cite{SouailleRoux_CPC01_WHAM} and the multi-state
Bennett acceptance ratio (MBAR) method \cite{ShirtsChodera_JCP08_MBAR}.
The most common use of reweighting is to obtain equilibrium expectations
or free energy differences. Reweighting can also be applied to obtain
dynamical information from the available contiguous trajectory pieces
\cite{SriramanKevrekidisHummer_JPCB109_6479}. When the probability
density of trajectories can be evaluated, MBAR can be applied to trajectories
instead of sample configurations, obtaining estimates of dynamical
expectations \cite{MinhChodera_JCP09_PathReweighting,ChoderaEtAl_JCP11_DynamicalReweighting,PrinzEtAl_JCP11_Reweighting,WuBolhuis_JCP13_MSTIS}.
Both WHAM and MBAR are statistically optimal under specific assumptions
as they can be derived from maximum-likelihood or minimum variance
principles \cite{FerrenbergSwendsen_PRL89_WHAM,Kumar1992,BartelsKarplus_JCC97_WHAM-ML,ShirtsChodera_JCP08_MBAR}.
However, a key assumption of both estimators is that data is given
as statistically independent samples of the respective equilibrium
distributions. In reality, MD and MCMC simulations provide time-correlated
data trajectories. Consequently, estimators exploiting the time-correlation
in the data can achieve significantly better results\cite{SakurabaKitao_JCTC09_MMMM,WuNoe_MMS14_TRAM1,MeyWuNoe_xTRAM,RostaHummer_DHAM}.
In particular, WHAM and MBAR cannot obtain unbiased estimates from
datasets for which the initial conditions of harvested trajectories
do not come from a probability density that is known a priori. Examples
of such datasets are swarms of short uncoupled trajectories \cite{ShirtsPande_Science2000_FoldingAtHome,NoeSchuetteReichWeikl_PNAS09_TPT,BuchFabritiis_PNAS11_Binding}
or Metadynamics and conformational flooding during the fill-up phase
\cite{LaioParrinello_PNAS99_12562,Grubmueller_PhysRevE52_2893}.

A complementary approach to address sampling problems is Markov state
modeling \cite{SchuetteFischerHuisingaDeuflhard_JCompPhys151_146,Swope2004,Singhal2005,SriramanKevrekidisHummer_JPCB109_6479,NoeHorenkeSchutteSmith_JCP07_Metastability,ChoderaEtAl_JCP07,BucheteHummer_JPCB08,PrinzEtAl_JCP10_MSM1}.
A Markov state model (MSM) transition matrix contains conditional
transition probabilities between discrete configuration states at
some lagtime $\tau$. Given the transition matrix, an equilibrium
distribution can be computed that is unbiased if each transition event
used for estimating the MSM originates from a local equilibrium distribution
restricted to respective discrete state \cite{SarichNoeSchuette_MMS09_MSMerror}.
In this case, MSMs are able to reweigh trajectories that are not sampled
from global equilibrium without any a priori knowledge of the transition
probabilities. In order to use MSMs for computing kinetics, $\tau$
must be long enough for transition events to approximately decorrelate
for a given configuration space discretization \cite{Swope2004,SarichNoeSchuette_MMS09_MSMerror,PrinzEtAl_JCP10_MSM1}.
Since MSM estimators are purely based on observed transition statistics
they cannot combine the information from different thermodynamic states.
Therefore, the orders-of-magnitude speedup that can sometimes be achieved
with enhanced sampling methods has not been accessible to MSMs as
yet.

The transition-based reweighting analysis method (TRAM) \cite{WuNoe_MMS14_TRAM1,MeyWuNoe_xTRAM}
aims at combining the advantages of reweighting estimators and MSMs.
In Ref. \cite{MeyWuNoe_xTRAM}, we have defined TRAM as a class of
estimators that (1) take the statistical weights of samples at different
thermodynamic states into account, in order to reweigh these samples;
and (2) exploits transitions observed in the sampled trajectories,
without assuming that these trajectories are sampled from equilibrium.
Ref. \cite{Tan_JStatPlanInf08_MonteCarloIntegration} introduced a
statistically optimal estimator for non-equilibrium trajectories given
that the statistical weight of each trajectory can be evaluated. Conceptually,
an optimal TRAM estimator could be formulated from this principle.
In practice, however, data is typically not stored at every integration
time step, such that trajectory probability densities are not available. 

In Ref. \cite{WuNoe_MMS14_TRAM1}, we have introduced the first TRAM
estimator that is applicable to practical molecular dynamics data,
and could show that it can provide superior estimates of equilibrium
probabilities and free energy compared with WHAM. However, the estimator
in Ref. \cite{WuNoe_MMS14_TRAM1} is only approximately optimal and
is very tedious to compute. Another TRAM estimator is presented in
Ref. \cite{RostaHummer_DHAM} and called the dynamic histogram analysis
method (DHAM). DHAM was shown to avoid systematic errors that may
occur when analyzing umbrella sampling with WHAM. DHAM uses a dynamical
model (diffusion along a reaction coordinate) to relate transition
matrices of simulations at different bias potentials. This assumption
is helpful when a diffusion model is appropriate and there is one
or a few slow reaction coordinates only (e.g. as in the case of umbrella
sampling). In this way it regularizes the solution and therefore makes
the estimates statistically more stable. In Ref. \cite{MeyWuNoe_xTRAM}
we introduced xTRAM, which does not assume a specific dynamical model
but only exploits the fact that samples can be reweighted in order
to couple different thermodynamic states. Moreover, xTRAM is as yet
the only TRAM estimator that avoids the discretization of reweighting
factors required in the other TRAM methods and as well as in WHAM.
xTRAM may provide substantial improvements compared to MBAR for time-correlated
data, is shown to be asymptotically exact, and is shown to converge
to MBAR when the data is not time correlated \cite{MeyWuNoe_xTRAM}.
However, xTRAM was not derived from a maximum-likelihood or minimum-variance
principle and is therefore probably not statistically optimal for
finite data sets.

In the present paper, we provide for the first time a statistically
optimal TRAM method by presenting a maximum-likelihood solution to
the discrete TRAM problem formulated in Ref. \cite{WuNoe_MMS14_TRAM1}.
We derive a set of self-consistent equations whose solution yields
the maximum likelihood dTRAM estimator. It is shown that the dTRAM
solution is an asymptotically correct estimator, i.e. it converges
to the exact equilibrium probabilities and transition probabilities
with an increasing amount of simulation data. The dTRAM equations
can be solved using a Newton method, as done before for WHAM and MBAR
\cite{TanLevy_JCP12_UWHAM,ZhuHummer_JCC12_ConvergenceWHAM}, or an
easy-to-implement iterative algorithm provided here.

We show that dTRAM becomes identical to WHAM in the limit of statistically
independent samples, i.e. sampled from the global equilibrium at each
thermodynamic state. Moreover, dTRAM becomes identical to a reversible
MSM when we have only a single thermodynamic state. A number of applications
are shown to demonstrate the usefulness and versatility of dTRAM.

\section{Discrete TRAM}

\subsection{Likelihood of WHAM, reversible Markov models and TRAM}

We assume that a set of MD or MCMC simulations have been performed,
each in one of $K$ thermodynamic states (indexed by the superscript
$k\in\{1,...,K\}$). For simulations in which the thermodynamic state
is frequently changed, such as in replica-exchange simulations, each
contiguous sequence is treated as a separate trajectory at one of
the $K$ thermodynamic states. Furthermore, we assume that the data
has been discretized to a configuration space partition (indexed by
subscripts $i,j\in\{1,...,n\}$). We are primarily interested in the
free energy, or equivalently, the equilibrium probability of discrete
states in some unbiased or reference ensemble $(\pi_{i})_{i=1,...,n}$.
In addition we might be interested in the equilibrium probability
of states under all biased ensembles. If the simulation trajectories
are long enough, we will also be able to compute kinetic properties,
as discussed later.

We will be dealing with simulations where the unbiased, or reference
probability $\pi_{i}$ and the biased probability at simulation condition
$k$, $\pi_{i}^{(k)}$ are related by a known and constant bias factor
$\gamma_{i}^{(k)}$:

\begin{eqnarray}
\pi_{i}^{(k)} & = & f^{(k)}\pi_{i}\gamma_{i}^{(k)},\label{eq:reweighting_1}\\
f^{(k)} & = & \frac{1}{\sum_{l}\pi_{l}\gamma_{l}^{(k)}}\label{eq:reweighting_2}
\end{eqnarray}
where $f^{(k)}$ is a normalization constant. Thus, the bias is multiplicative
in probabilities or additive in the potential. This formalism is applicable
whenever one has simulations conducted at different thermodynamic
states, such as replica-exchange methods or umbrella sampling, but
also direct simulations at different temperatures or Hamiltonians.
In the results section, we will show how the reweighting factor $\gamma_{i}^{(k)}$
can be computed for a few selected examples. The most common analysis
method used in the present scenario is WHAM. WHAM uses the histogram
counts $N_{i}^{(k)}$, i.e. the number of samples falling into bin
$i$ at thermodynamic state $k$. Although WHAM was originally derived
as a minimum-variance estimator\cite{FerrenbergSwendsen_PRL89_WHAM,Kumar1992},
it can be derived as a maximum-likelihood estimator\cite{BartelsKarplus_JCC97_WHAM-ML}
using the likelihood
\begin{equation}
L_{\mathrm{WHAM}}=\prod_{k}\prod_{i}(\pi_{i}^{(k)})^{N_{i}^{(k)}}\label{eq:WHAM-likelihood}
\end{equation}
which simply assumes that every count $N_{i}^{(k)}$ is independently
drawn from the biased distribution $\pi_{i}^{(k)}$, which is linked
to the unbiased distribution $\boldsymbol{\pi}$ \emph{via} Eq. (\ref{eq:reweighting_1}).

Let us now turn to reversible Markov state models \cite{Noe_JCP08_TSampling,BucheteHummer_JPCB08,Bowman_JCP09_Villin,PrinzEtAl_JCP10_MSM1}.
The maximum likelihood Markov model is the transition matrix $\mathbf{P}=(p_{ij})$
between $n$ discrete configuration states, that maximizes the likelihood
of the observed transitions between these states. The likelihood of
a Markov model is well known\cite{Book_Norris}, and simply a product
of all transition probabilities corresponding to the observed trajectory.
To obtain a reversible Markov state model, this likelihood is maximized
using the constraints that the transition probabilities $p_{ij}$
must fulfill detailed balance with respect to the equilibrium distribution
$\boldsymbol{\pi}$:
\begin{eqnarray}
L_{\mathrm{MSM}} & = & \prod_{i}\prod_{j}p_{ij}^{c_{ij}}\label{eq:MSM-likelihood}\\
s.t. &  & \pi_{i}p_{ij}=\pi_{j}p_{ji}\:\:\:\:\:\mathrm{for}\:\mathrm{all}\: i,j,\label{eq:MSM-DBconstraint}
\end{eqnarray}
where $c_{ij}$ is the number of times the trajectories were observed
in state $i$ at time $t$ and in state $j$ at a later time $t+\tau$,
where $\tau$ is the lag time at which the Markov model is estimated.
For an MSM, all simulation data producing counts $c_{ij}$, has to
be generated at the same thermodynamic state (e.g. temperature, Hamiltonian),
and the estimated $\mathbf{P}$ is then only valid for this thermodynamic
state. The reversibility of the MSM is ensured by the constraint equations
(\ref{eq:MSM-DBconstraint}). Estimators that maximize Eqs. (\ref{eq:MSM-likelihood}-\ref{eq:MSM-DBconstraint})
usually provide both $\mathbf{P}$ and the equilibrium distribution
$\boldsymbol{\pi}$ \cite{Bowman_JCP09_Villin,PrinzEtAl_JCP10_MSM1}.

In TRAM, we combine these two approaches as follows: we avoid the
WHAM assumption that every count is sampled from global equilibrium,
and instead treat every trajectory at thermodynamic condition $k$
as a Markov chain with the configuration-state transition counts $c_{ij}^{(k)}$.
However, in contrast to Markov models we exploit the fact that equilibrium
probabilities can be reweighted between different thermodynamic states
via (\ref{eq:reweighting_1}-\ref{eq:reweighting_2}). The resulting
likelihood of all $\mathbf{P}^{(k)}$ and $\boldsymbol{\pi}$, based
on simulations at all thermodynamic states can be formulated as: 
\begin{eqnarray}
L_{\mathrm{TRAM}} & = & \prod_{k}\prod_{i}\prod_{j}(p_{ij}^{(k)})^{c_{ij}^{(k)}}\label{eq:TRAM-likelihood}\\
s.t. &  & \pi_{i}^{(k)}p_{ij}^{(k)}=\pi_{j}^{(k)}p_{ji}^{(k)}\:\:\:\:\:\mathrm{for}\:\mathrm{all}\: i,j,k.\label{eq:TRAM-DBconstraint}
\end{eqnarray}
Here, $\mathbf{P}^{(k)}=(p_{ij}^{(k)})$ is the Markov transition
matrix at thermodynamic state $k$, and $c_{ij}^{(k)}$ are the number
of transitions observed at that simulation condition. $\boldsymbol{\pi}^{(k)}$
is the vector of equilibrium probabilities of discrete states at each
thermodynamic state. Note that all of these $K$ equilibrium distributions
are coupled through Eqs. (\ref{eq:reweighting_1}-\ref{eq:reweighting_2}).
Because each Markov model $\mathbf{P}^{(k)}$ must have the distribution
$\boldsymbol{\pi}^{(k)}$ as a stationary distribution, all Markov
models are coupled too. This is what makes the maximization of the
TRAM likelihood Eqs. (\ref{eq:TRAM-likelihood}-\ref{eq:TRAM-DBconstraint})
difficult, and it can neither be achieved by WHAM, nor by existing
MSM estimators. We call Eqs. (\ref{eq:reweighting_1},\ref{eq:reweighting_2},\ref{eq:TRAM-likelihood},\ref{eq:TRAM-DBconstraint})
the TRAM equations. In the present paper we will obtain the reweighting
factors $\gamma_{i}^{(k)}$ in Eqs. (\ref{eq:reweighting_1}-\ref{eq:reweighting_2})
by a configuration space discretization or binning, such as in WHAM.
For this reason we call the present solution a discrete TRAM method,
which should be distinguished from approaches where the reweighting
is done for individual samples \cite{ShirtsChodera_JCP08_MBAR,MeyWuNoe_xTRAM}.

\subsection{dTRAM log-likelihood and self-consistent solution equations}

We will seek the maximum likelihood of Eq. (\ref{eq:TRAM-likelihood}).
As in common practice, we work with the logarithm of the likelihood,
because it has the same maximum point as the likelihood but can be
treated more easily:
\begin{equation}
\log L_{\mathrm{TRAM}}=\sum_{k=1}^{K}\sum_{i=1}^{n}\sum_{j=1}^{n}c_{ij}^{(k)}\ln p_{ij}^{(k)}\label{eq:TRAM-log-likelihood}
\end{equation}
Moreover, we have the following constraints. Using detailed balance
Eq. (\ref{eq:TRAM-DBconstraint}) with the reweighting equations (\ref{eq:reweighting_1}-\ref{eq:reweighting_2})
results in
\begin{equation}
\pi_{i}\gamma_{i}^{(k)}p_{ij}^{(k)}=\pi_{j}\gamma_{j}^{(k)}p_{ji}^{(k)}\:\:\:\:\mathrm{for\: all\:}i,j,k.\label{eq:TRAM-constraint1}
\end{equation}
Note that the normalization factors, $f^{(k)}$, have cancelled. In
addition, $\mathbf{P}^{(k)}$ should be a transition matrix and $\boldsymbol{\pi}$
should be a probability vector, so we have the normalization conditions:

\begin{eqnarray}
\sum_{j}p_{ij}^{(k)} & = & 1\:\:\:\:\forall i,k\label{eq:TRAM-constraint2}\\
\sum_{j}\pi_{j} & = & 1\label{eq:TRAM-constraint3}
\end{eqnarray}
The normalization of $\boldsymbol{\pi}^{(k)}$ is achieved by the
normalization constants in Eq. (\ref{eq:reweighting_1})-(\ref{eq:reweighting_2}).

In order to solve the discrete TRAM problem we have to maximize the
log likelihood (\ref{eq:TRAM-log-likelihood}) under the constraints
(\ref{eq:TRAM-constraint1}-\ref{eq:TRAM-constraint3}). The variables
are both the unbiased equilibrium probabilities $\boldsymbol{\pi}$
(providing $n-1$ variables due to the constraint (\ref{eq:TRAM_iteration_pi})),
and the biased transition matrices $\mathbf{P}^{(k)}$ (each having
$n(n-1)/2$ remaining free variables that are not fixed by constraints
(\ref{eq:TRAM-constraint1})-(\ref{eq:TRAM-constraint2})). 

Note that changing the simulation conditions, such as bias or temperature,
will modify the transition probabilities in a non-trivial way that
depends on the simulation condition, the integrator and thermostat
used, and the state space discretization. Therefore we cannot relate
the different $\mathbf{P}^{(k)}$ without restricting the generality
of our estimator. The only general connection between these Markov
models is the coupling of their equilibrium distributions via Eqs.
(\ref{eq:reweighting_1}-\ref{eq:reweighting_2}). 

In Appendix A.1-3, we use Lagrange duality theory to show that the
optimal solution of the discrete TRAM problem above fulfills the following
two conditions:
\begin{eqnarray}
\sum_{k}\sum_{j}\frac{\left(c_{ij}^{(k)}+c_{ji}^{(k)}\right)\gamma_{i}^{(k)}\pi_{i}v_{j}^{(k)}}{\gamma_{i}^{(k)}\pi_{i}v_{j}^{(k)}+\gamma_{j}^{(k)}\pi_{j}v_{i}^{(k)}} & = & \sum_{k}\sum_{j}c_{ji}^{(k)}\label{eq:dTRAM_equation1}\\
\sum_{j}\frac{\left(c_{ij}^{(k)}+c_{ji}^{(k)}\right)\gamma_{j}^{(k)}\pi_{j}}{\gamma_{i}^{(k)}\pi_{i}v_{j}^{(k)}+\gamma_{j}^{(k)}\pi_{j}v_{i}^{(k)}} & = & 1\label{eq:dTRAM_equation2}
\end{eqnarray}

where $v_{i}^{(k)}$ are unknown Lagrange multipliers. In the setting
with detailed balance we can unfortunately not give a closed expression
for them, but we can optimize them along with the equilibrium distribution
$\boldsymbol{\pi}$. Note that the equations above do not require
the transition probabilities $p_{ij}^{(k)}$ to be computed explicitly.
If these are desired, they can be subsequently computed from the solution
of Eqs. (\ref{eq:dTRAM_equation1}-\ref{eq:dTRAM_equation2}) (see
Sec. \ref{sub:kinetics} below).

In Appendix A.4 we prove that the dTRAM equations above are asymptotically
correct. This means that in the limit of a lot of simulation data
- either realized by long trajectories or many short trajectories
- the estimate will converge to the correct stationary distributions
$\boldsymbol{\pi}$ and $\boldsymbol{\pi}^{(k)}$.

\subsection{Solution methods}

We can rewrite the self-consistent equations (\ref{eq:dTRAM_equation1},\ref{eq:dTRAM_equation2})
to derive the following iteration (fixed-point method), that can be
used to numerically solve the discrete TRAM problem. First we initialize
$\boldsymbol{\pi}$ and $\mathbf{v}^{(k)}$ by the simple guess:
\begin{eqnarray}
\pi_{i}^{init} & := & 1/n\label{eq:initial_guess_pi}\\
v_{i}^{(k),init} & := & \sum_{j}c_{ij}^{(k)}
\end{eqnarray}
and then we iterate the following equations until $\boldsymbol{\pi}$
is converged:
\begin{eqnarray}
v_{i}^{(k),new} & := & v_{i}^{(k)}\sum_{j}\frac{\left(c_{ij}^{(k)}+c_{ji}^{(k)}\right)\gamma_{j}^{(k)}\pi_{j}}{\gamma_{i}^{(k)}\pi_{i}v_{j}^{(k)}+\gamma_{j}^{(k)}\pi_{j}v_{i}^{(k)}}\label{eq:TRAM_iteration_v}\\
\pi_{i}^{new} & := & \frac{\sum_{k,j}c_{ji}^{(k)}}{\sum_{k,j}\frac{\left(c_{ij}^{(k)}+c_{ji}^{(k)}\right)\gamma_{i}^{(k)}v_{j}^{(k)}}{\gamma_{i}^{(k)}\pi_{i}v_{j}^{(k)}+\gamma_{j}^{(k)}\pi_{j}v_{i}^{(k)}}}\label{eq:TRAM_iteration_pi}
\end{eqnarray}
Instead of the simple $1/n$ initialization for $\pi_{i}$ in Eq.
(\ref{eq:initial_guess_pi}), we could use the standard WHAM algorithm
to obtain a much better guess \cite{FerrenbergSwendsen_PRL89_WHAM,Kumar1992}.
While a better starting point might be relevant for optimizing computational
performance, we have not observed the estimation result to depend
on this choice. 

As an alternative to the fixed-point iteration (\ref{eq:TRAM_iteration_v},\ref{eq:TRAM_iteration_pi}),
we can solve equations (\ref{eq:dTRAM_equation1},\ref{eq:dTRAM_equation2})
by using the multidimensional Newton method for root finding available
in many numerics packages.

\subsection{Kinetics and the selection of the estimation lag time $\tau$}

\label{sub:kinetics}

Given $\boldsymbol{\pi}$ and $\mathbf{v}^{(k)}$ at their optimal
values, the transition probabilities can be computed for any thermodynamic
state $k$ simulated at by:
\begin{equation}
p_{ij}^{(k)}(\tau)=\frac{\left(c_{ij}^{(k)}(\tau)+c_{ji}^{(k)}(\tau)\right)\gamma_{j}^{(k)}\pi_{j}}{\gamma_{i}^{(k)}\pi_{i}v_{j}^{(k)}+\gamma_{j}^{(k)}\pi_{j}v_{i}^{(k)}}\label{eq:pij}
\end{equation}
See Appendix A.2 for the derivation. In Eq. (\ref{eq:pij}) we have
explicitly stated that transition counts, and hence the transition
probabilities are estimated at a given lag time $\tau$. As a consequence
of the asymptotic correctness of dTRAM (Appendix A.4), the estimates
of $p_{ij}^{(k)}(\tau)$ are also asymptotically correct, that is
for either long trajectories or many short trajectories we will get
an unbiased estimate of the transition probabilities.

In order to compute kinetics, such as transition rates or timescales,
the transition matrices $\mathbf{P}^{(k)}$ do not only have to be
valid for the lag time $\tau$ estimated at, but they have to be Markov
models that predict the kinetics at longer times correctly. How adequate
$\mathbf{P}^{(k)}$ is as a Markov model should be tested by validating
that the relaxation timescales computed from the eigenvalues of $\mathbf{P}^{(k)}$
are approximately constant in $\tau$ \cite{Swope2004} and by checking
that the Chapman-Kolmogorow $\mathbf{P}^{(k)}(n\tau)\approx[\mathbf{P}^{(k)}(\tau)]^{n}$
approximately holds \cite{PrinzEtAl_JCP10_MSM1}.

The $\mathbf{P}^{(k)}$ can only be used as Markov models when the
contiguous simulation trajectories are long enough to support a suitable
lag time $\tau$. Generalized ensemble simulations, such as replica-exchange,
parallel or simulated tempering generally only provide very short
contiguous trajectory pieces and are only suitable for constructing
Markov models of small systems and using excellent configuration state
discretizations \cite{SriramanKevrekidisHummer_JPCB109_6479,ChoderaEtAl_JCP11_DynamicalReweighting,PrinzEtAl_JCP11_Reweighting}. 

Based on umbrella sampling simulations, the construction of Markov
models at the different umbrellas $k$ is usually possible, but for
the unbiased system we can only obtain the equilibrium distribution
$\boldsymbol{\pi}$ and not the Markov model. The reason is that the
transition matrices (\ref{eq:pij}) can only be estimated at the different
simulation conditions $k$, whereas the equilibrium probability of
the chosen reference ensemble $\boldsymbol{\pi}$ is computed through
reweighting, and is thus also available for thermodynamic states not
simulated at. However, the umbrella-Markov models $\mathbf{P}^{(k)}$
could still provide useful information. For example comparing the
longest relaxation timescale of each umbrella Markov model with the
respective simulation length could be used as an indicator of convergence,
and whether some or all simulation lengths should be increased \cite{RostaHummer_DHAM}. 

Unbiased MD simulations at different thermodynamic states are most
suitable for constructing Markov models, because one has the choice
of running simulations long enough to accommodate a suitable lag time
$\tau$. A systematic way of constructing such simulations is the
random swapping protocol \cite{MeyWuNoe_xTRAM}. Note that such simulations
may not only violate the sampling from global equilibrium, but also
the sampling from local equilibrium, it is possible that the estimation
of $\boldsymbol{\pi}$ and all associated stationary estimates are
biased for short lag times $\tau$. Therefore, when using dTRAM to
analyze unbiased MD simulations at different thermodynamic states,
one should definitely compute the estimates as a function of $\tau$
in order to ensure that a large enough $\tau$ is used to obtain unbiased
estimates.

\subsection{WHAM is a special case of dTRAM}

We now show that the commonly used WHAM method is obtained as a special
case of dTRAM. Starting from the dTRAM estimator, we employ the WHAM
assumption that each sample at thermodynamic state $k$ is independently
generated from the biased probability distribution $\pi^{(k)}$. This
means that transition probabilities $p_{ij}^{(k)}$ are equal to the
probability of observing state $j$ without knowledge of $i$:
\begin{equation}
p_{ij}^{(k)}=\pi_{j}^{(k)}\label{eq:p-ind}
\end{equation}
In a setting where counts are generated independently, the transition
counts $c_{ij}^{(k)}$ can be modeled by splitting up the total counts
ending in bin $j$ according to the equilibrium probability that they
have been in a given bin $i$ before:
\begin{equation}
c_{ij}^{(k)}=\pi_{i}^{(k)}N_{j}^{(k)}\label{eq:c-ind}
\end{equation}
Note that this selection generates actually observed histogram counts
as $\sum_{i}c_{ij}^{(k)}=N_{j}^{(k)}\sum_{i}\pi_{i}^{(k)}=N_{j}^{(k)}$.
Substituting $\pi_{j}^{(k)}$ in (\ref{eq:p-ind}-\ref{eq:c-ind})
using (\ref{eq:reweighting_1}-\ref{eq:reweighting_2}) and inserting
the result into Eq. (\ref{eq:pij}) yields the equalities
\begin{equation}
N_{j}^{(k)}\gamma_{i}^{(k)}\pi_{i}+N_{i}^{(k)}\gamma_{j}^{(k)}\pi_{j}=\gamma_{i}^{(k)}\pi_{i}v_{j}^{(k)}+\gamma_{j}^{(k)}\pi_{j}v_{i}^{(k)}
\end{equation}
which must hold for all $i$ and $k$. This is exactly the case when
the Lagrange multipliers become:
\begin{equation}
v_{i}^{(k)}=N_{i}^{(k)}.\label{eq:nu-c}
\end{equation}
Substituting (\ref{eq:c-ind}) and (\ref{eq:nu-c}) into (\ref{eq:TRAM_iteration_pi})
gives us the solution for the unbiased stationary probabilities:
\begin{eqnarray}
\pi_{i}^{\mathrm{new}} & = & \frac{\sum_{k}N_{i}^{(k)}}{\sum_{k}N^{(k)}f^{(k)}\gamma_{i}^{(k)}}\label{eq:WHAM_iteration_pi}\\
f^{(k),\mathrm{new}} & = & \frac{1}{\sum_{j}\gamma_{j}^{(k)}\pi_{j}}\label{eq:WHAM_iteration_f}
\end{eqnarray}
which is exactly the WHAM algorithm \cite{FerrenbergSwendsen_PRL89_WHAM,Kumar1992}.
Therefore, WHAM is a special case of dTRAM, suggesting that TRAM should
yield estimates that are at least as good as WHAM, but should give
better estimates when the WHAM assumptions of sampling from global
equilibrium at condition $k$ does not hold.

\subsection{A reversible Markov state model is a special case of dTRAM}

Now we relate dTRAM to reversible Markov models. Suppose we only have
a single thermodynamic state $k$ and one or several simulation trajectories
generating counts $c_{ij}$ at this condition. In this case we can
drop the index $k$, all reweighting factors are unity $\gamma_{i}\equiv1$,
and equations (\ref{eq:dTRAM_equation1}-\ref{eq:dTRAM_equation2})
become:
\begin{eqnarray}
\sum_{j}\frac{\left(c_{ij}+c_{ji}\right)\pi_{i}v_{j}}{\pi_{i}v_{j}+\pi_{j}v_{i}} & = & \sum_{j}c_{ji}\label{eq:dTRAM_equation1-1}\\
\sum_{j}\frac{\left(c_{ij}+c_{ji}\right)\pi_{j}}{\pi_{i}v_{j}+\pi_{j}v_{i}} & = & 1\label{eq:dTRAM_equation2-1}
\end{eqnarray}
We can combine both equations to:
\begin{eqnarray}
\sum_{j}c_{ji}+v_{i} & = & \sum_{j}\frac{\left(c_{ij}+c_{ji}\right)\pi_{i}v_{j}}{\pi_{i}v_{j}+\pi_{j}v_{i}}+\sum_{j}\frac{\left(c_{ij}+c_{ji}\right)\pi_{j}v_{i}}{\pi_{i}v_{j}+\pi_{j}v_{i}}\nonumber \\
 & = & \sum_{j}\left(c_{ij}+c_{ji}\right)
\end{eqnarray}
Thus we solve for the Lagrange multipliers:
\begin{equation}
v_{i}=\sum_{j}c_{ij}=N_{i}.
\end{equation}
Substituting $v_{i}=N_{i}$ into (\ref{eq:dTRAM_equation2}) leads
to the optimality condition for $\pi_{i}$:
\begin{eqnarray}
\pi_{i} & = & \sum_{j}\frac{c_{ij}+c_{ji}}{\frac{c_{i}}{\pi_{i}}+\frac{c_{j}}{\pi_{j}}}
\end{eqnarray}
Inserting the result into (\ref{eq:pij}) yields the reversible transition
matrix estimator:
\begin{equation}
\pi_{i}p_{ij}=\frac{c_{ij}+c_{ji}}{\frac{c_{j}}{\pi_{j}}+\frac{c_{i}}{\pi_{i}}}
\end{equation}
which is identical to the known optimality condition for a reversible
Markov model transition matrix and the corresponding iterative estimator
\cite{Bowman_JCP09_Villin,PrinzEtAl_JCP10_MSM1}. Therefore, a reversible
MSM is a special case of dTRAM.

\section{Results}

\subsection{Illustrative example}

\begin{figure}[H]
\noindent \begin{centering}
(a)\includegraphics[width=0.95\columnwidth]{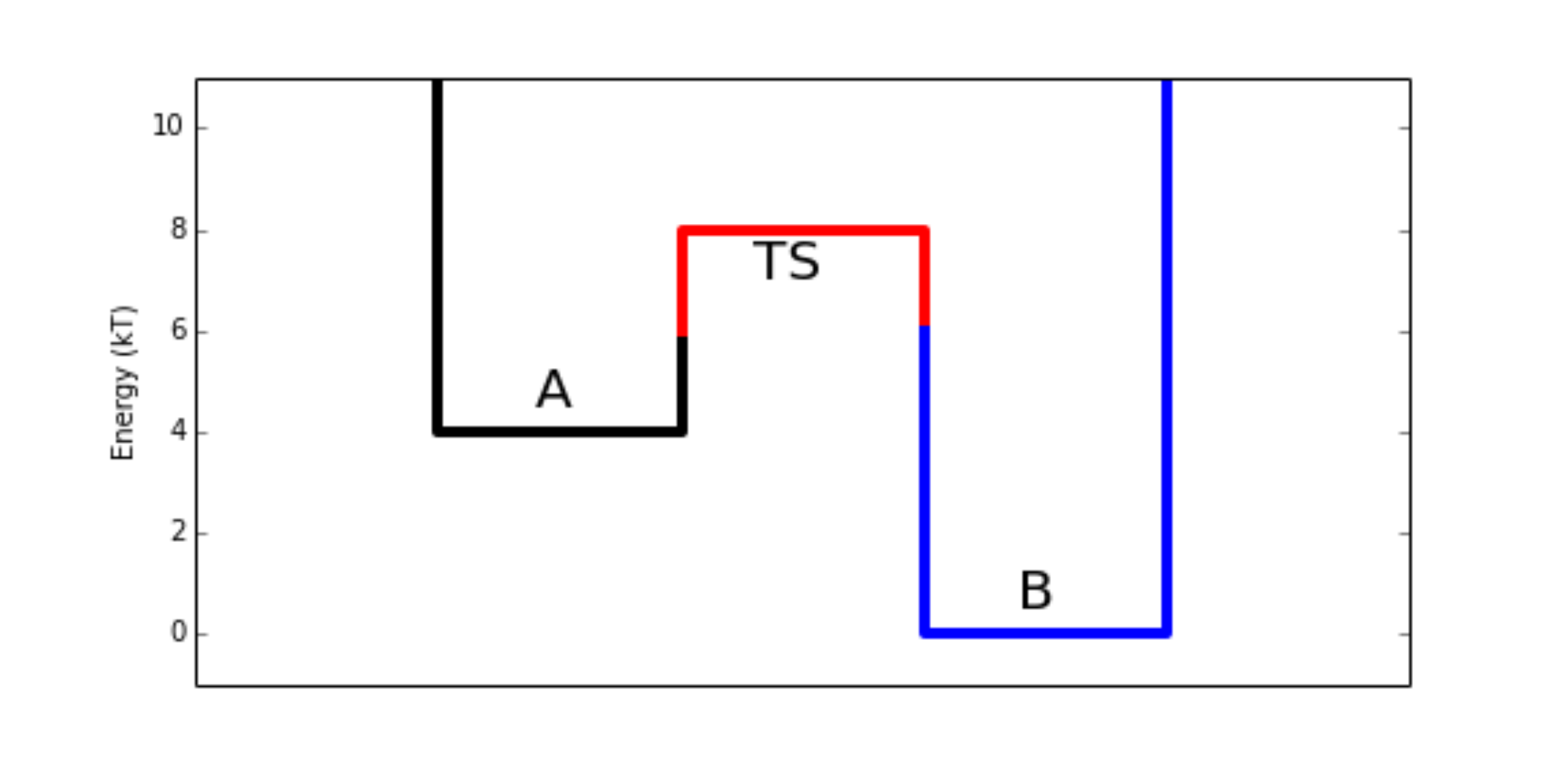}
\par\end{centering}

\noindent \begin{centering}
(b)\includegraphics[width=0.95\columnwidth]{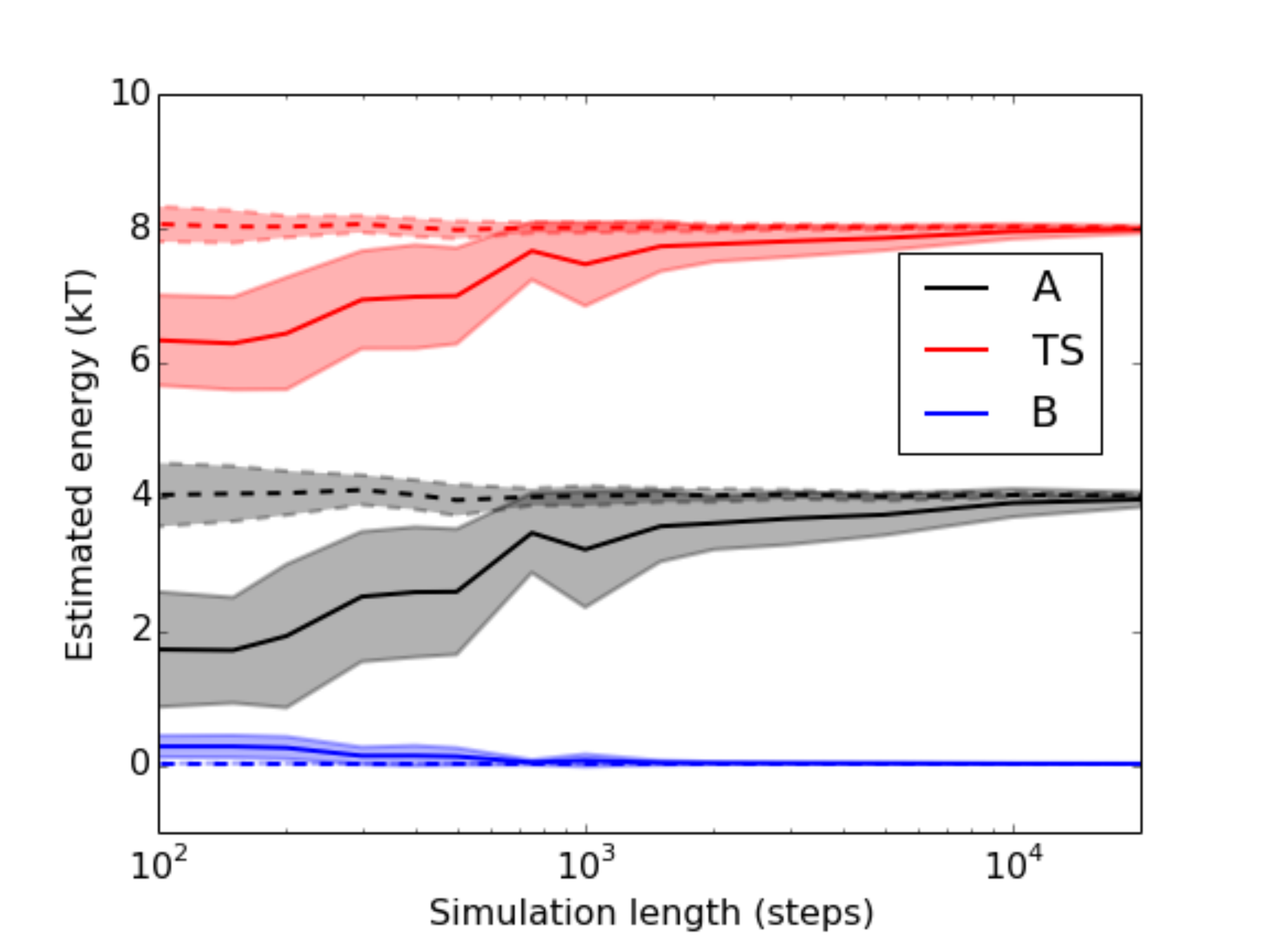}
\par\end{centering}

\protect\caption{\label{fig:discrete_example}Comparison of WHAM and dTRAM estimation
results using three independent trajectories sampling from three discrete
states (two unbiased trajectories starting in $A$ and $B$ respectively,
and biased trajectory sampling from a flat energy landscape over $A$,
$TS$ and $B$). (a) Schematic of the energies of three states. Error
bars correspond to one standard deviation of the estimate computed
over 25 independent runs. (b) Estimation of the energies of $A$,
$TS$ and $B$ using WHAM (solid lines) and dTRAM (dashed lines).}

\end{figure}

We start with a simple example to illustrate a scenario in which the
classical WHAM estimator fails because the assumption of sampling
from global equilibrium is not fulfilled. Fig. \ref{fig:discrete_example}a
shows the energy levels $(u_{A}=4,\: u_{TS}=8,\: u_{B}=0)$ of a discrete
three-state system. We consider a Metropolis-Hastings jump process
between neighboring states. Due to the high-energy transition state
$TS$, the minima $A$ and $B$ are long-lived and escaping them is
a rare event. Now we run a simulation consisting of three independent
trajectories:
\begin{enumerate}
\item An unbiased trajectory of length $L$ starting state $A$
\item An unbiased trajectory of length $L$ starting state $B$
\item A trajectory of length $L$ using bias $(b_{A}=4,\: b_{TS}=0,b_{B}=8)$
starting in state $TS$.
\end{enumerate}
The biased trajectory 3 samples from an energy landscape that is flat
over $A$, $B$ and $TS$. Trajectories 1 and 2 will be stuck in states
$A$ or $B$, respectively, and only be able to escape them when $L$
is sufficiently long. 

Trajectories 1 and 2 are using the same unbiased Hamiltonian and are
therefore in the same thermodynamic state $k=1$. The corresponding
reweighting factor is simply:
\[
\gamma_{i}^{(1)}=1
\]
for all states $i=A,\, TS,\, B$. Trajectory 3 is in a different thermodynamic
state with a biased Hamiltonian that we shall call $k=2$. We use
the bias as a reweighting factor:
\[
\gamma_{i}^{(2)}=\mathrm{e}^{-b_{i}}
\]
for all states $i=A,\, TS,\, B$. In this way we can reweigh all samples
between the two thermodynamic states and produce estimates of the
energies $u_{A},\, u_{TS},u_{B}$ using both WHAM with Eqs. (\ref{eq:WHAM_iteration_pi}-\ref{eq:WHAM_iteration_f})
and dTRAM with Eqs (\ref{eq:TRAM_iteration_v}-\ref{eq:TRAM_iteration_pi}).

Because states $A$ and $B$ are long-lived, we expect that the unbiased
trajectories 1 and 2 cannot sample from the global equilibrium, unless
the simulation times $L$ is very long. As a result the WHAM estimator
is strongly biased initially and systematically underestimates the
energy differences - see solid lines in Fig. \ref{fig:discrete_example}b.
Only after about $L=10,000$ steps, the bias of the WHAM estimator
is negligible. In contrast, the dTRAM estimate is unbiased even for
short simulation lengths - see dashed lines in Fig. \ref{fig:discrete_example}b.
dTRAM does not suffer from the WHAM bias because each transition count
is indeed independent from the previous transition count, even when
the simulation does not sample from global equilibrium. Moreover,
the uncertainty of the dTRAM estimate is much smaller than the uncertainty
of the WHAM estimate. This is because every transition count is an
independent sample, and therefore dTRAM benefits from a much larger
statistical efficiency than WHAM.

While illustrative, this example is over-simplistic. In reality, we
do not have discrete states and the dynamics is no Markov chain. Let
us investigate next how dTRAM performs in a more realistic setting.

\subsection{Umbrella sampling}

In umbrella sampling, one introduces for each simulation $k$ an additive
bias potential, such that our total potential in simulation $k$ is
given by
\begin{equation}
u_{i}^{(k)}=u_{i}+b_{i}^{(k)}
\end{equation}
where $i$ is a bin index of one or several finely discretized coordinates
in which the bias potential is acting. $u_{i}$ is the unbiased potential
evaluated at bin $i$ (usually at its center), and $b_{i}^{(k)}$
is the value of the $k$th umbrella (bias) potential. All energies
are dimensionless (i.e. in units of the thermal energy $k_{B}T$).
The biased equilibrium probabilities are proportional to:
\begin{eqnarray}
\pi_{i}^{(k)} & \propto & \mathrm{e}^{-u_{i}^{(k)}}=\gamma_{i}^{(k)}\pi_{i}
\end{eqnarray}
with the unbiased equilibrium probability and the reweighting factors
given by :
\begin{eqnarray}
\pi_{i} & = & \frac{\mathrm{e}^{-u_{i}}}{\sum_{j}\mathrm{e}^{-u_{j}}}\\
\gamma_{i}^{(k)} & = & \mathrm{e}^{-b_{i}^{(k)}}\label{eq:reweighting-factor-US}
\end{eqnarray}
Umbrella sampling is typically employed using stiff constraining potentials,
such that the simulations quickly decorrelate. In such a scenario,
WHAM is a suitable estimator for extracting unbiased equilibrium probabilities
and free energy differences. However, an analysis of umbrella sampling
data can indeed benefit from using dTRAM. Firstly, dTRAM uses conditional
transition events, and the number of statistically independent observations
thus depends on a lag time $\tau$ which can be thought of as a local
decorrelation time, i.e. a time at which transitions become statistically
independent. This time is generally shorter than the global decorrelation
time (or statistical inefficiency) $t_{corr}$, at which samples generated
by the trajectory become statistically independent. Thus, whenever
$t_{corr}$ is longer than the interval at which configurations are
saved, dTRAM will be able to exploit that this data more efficiently
than WHAM. This should lead to improved estimates. 

Moreover, umbrella sampling can, in conjunction with WHAM, result
in systematic errors that can be avoided with a TRAM estimator: As
shown in\cite{RostaHummer_DHAM}, using umbrellas that are too weak
to stabilize the simulation at a transition state can lead to umbrella
simulations that still have high internal free energy barriers and
therefore do not generate samples that are drawn from the respective
equilibrium distribution. Although a WHAM estimate would be correct
in the limit of infinitely long simulation times, it may provide drastically
biased estimates for practical simulation times.

\begin{figure}[H]
\noindent \begin{centering}
(A)\includegraphics[width=0.7\columnwidth]{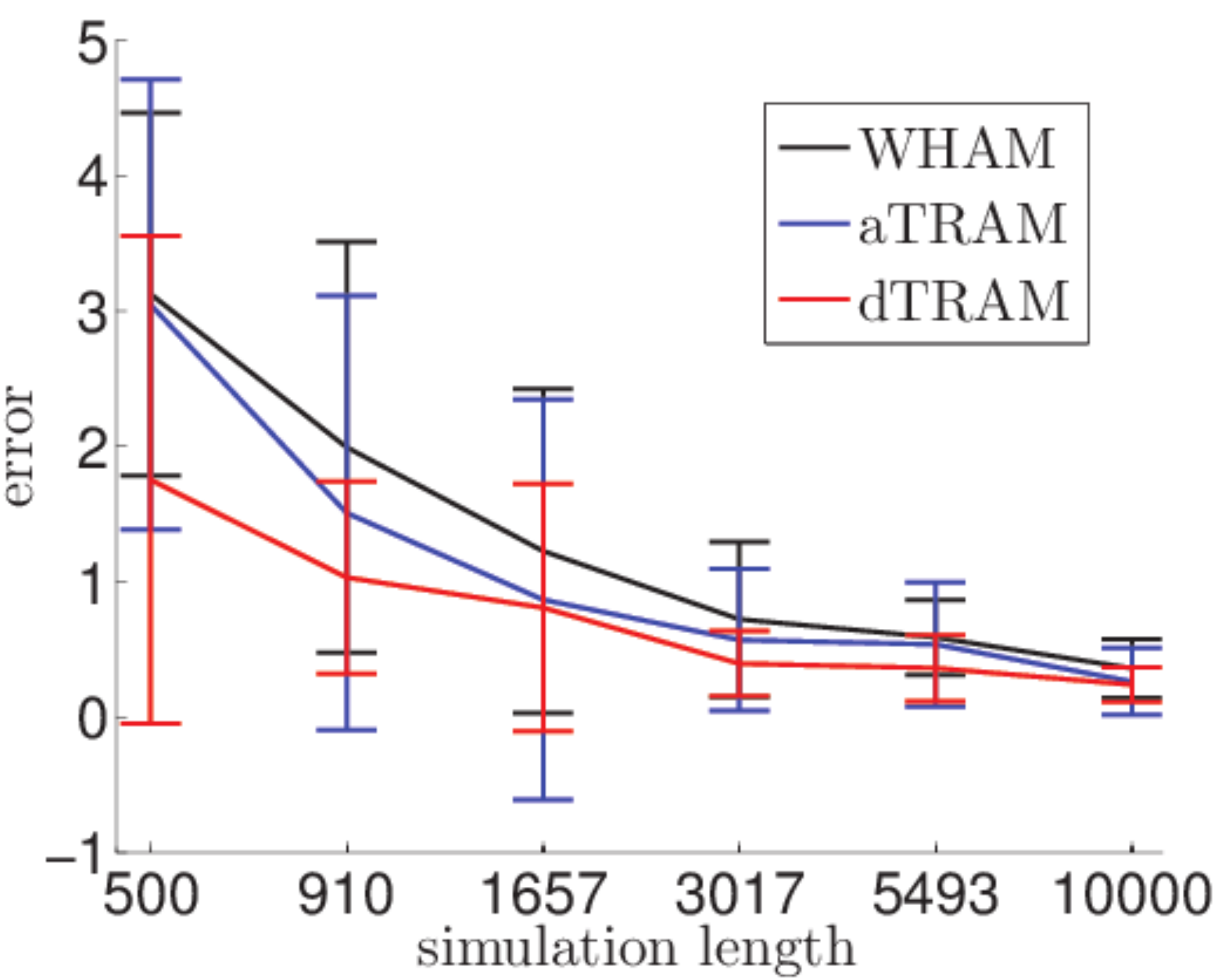} 
\par\end{centering}

\noindent \begin{centering}
(B) \includegraphics[width=0.7\columnwidth]{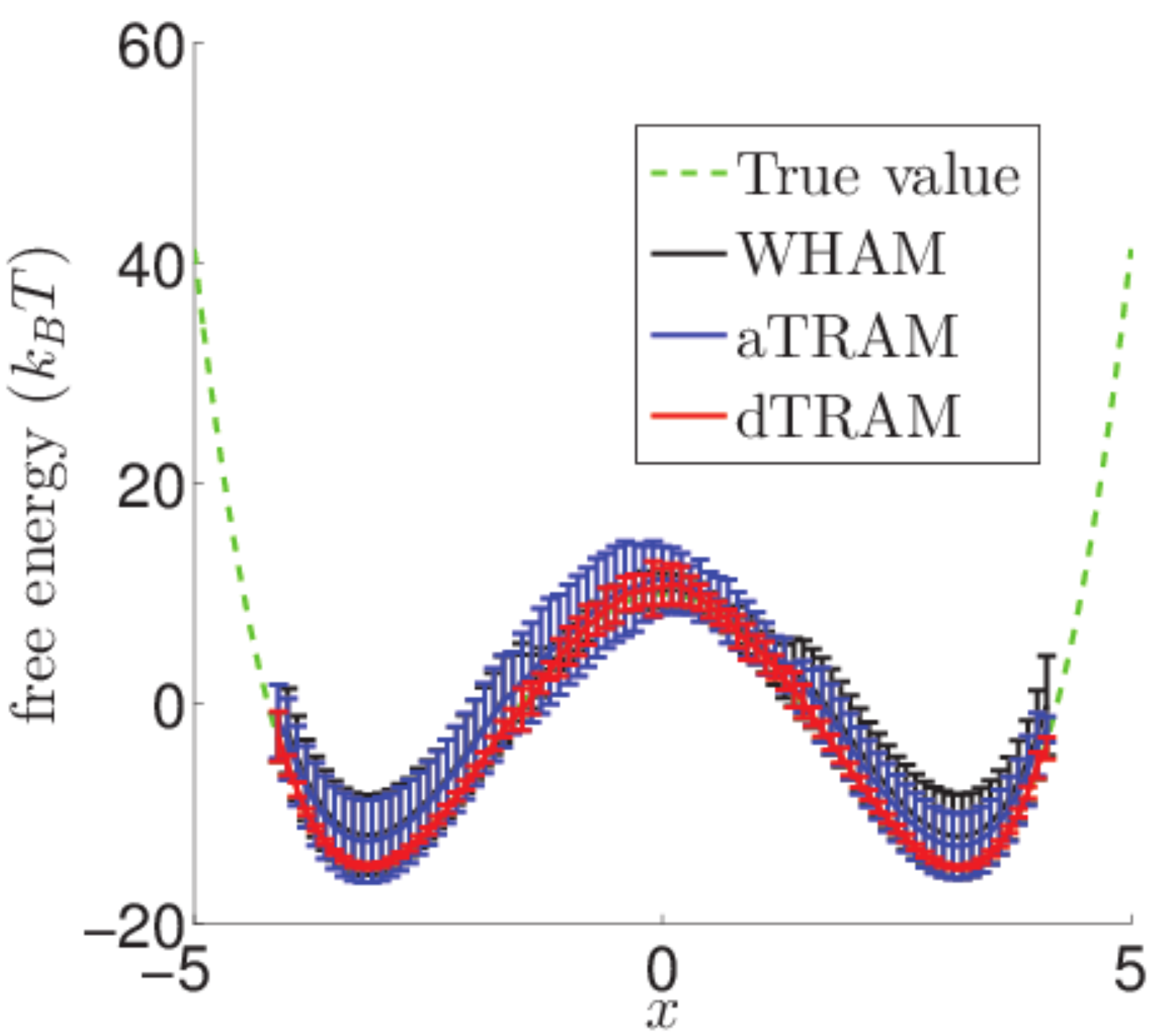}
\par\end{centering}

\noindent \begin{centering}
(C) \includegraphics[width=0.7\columnwidth]{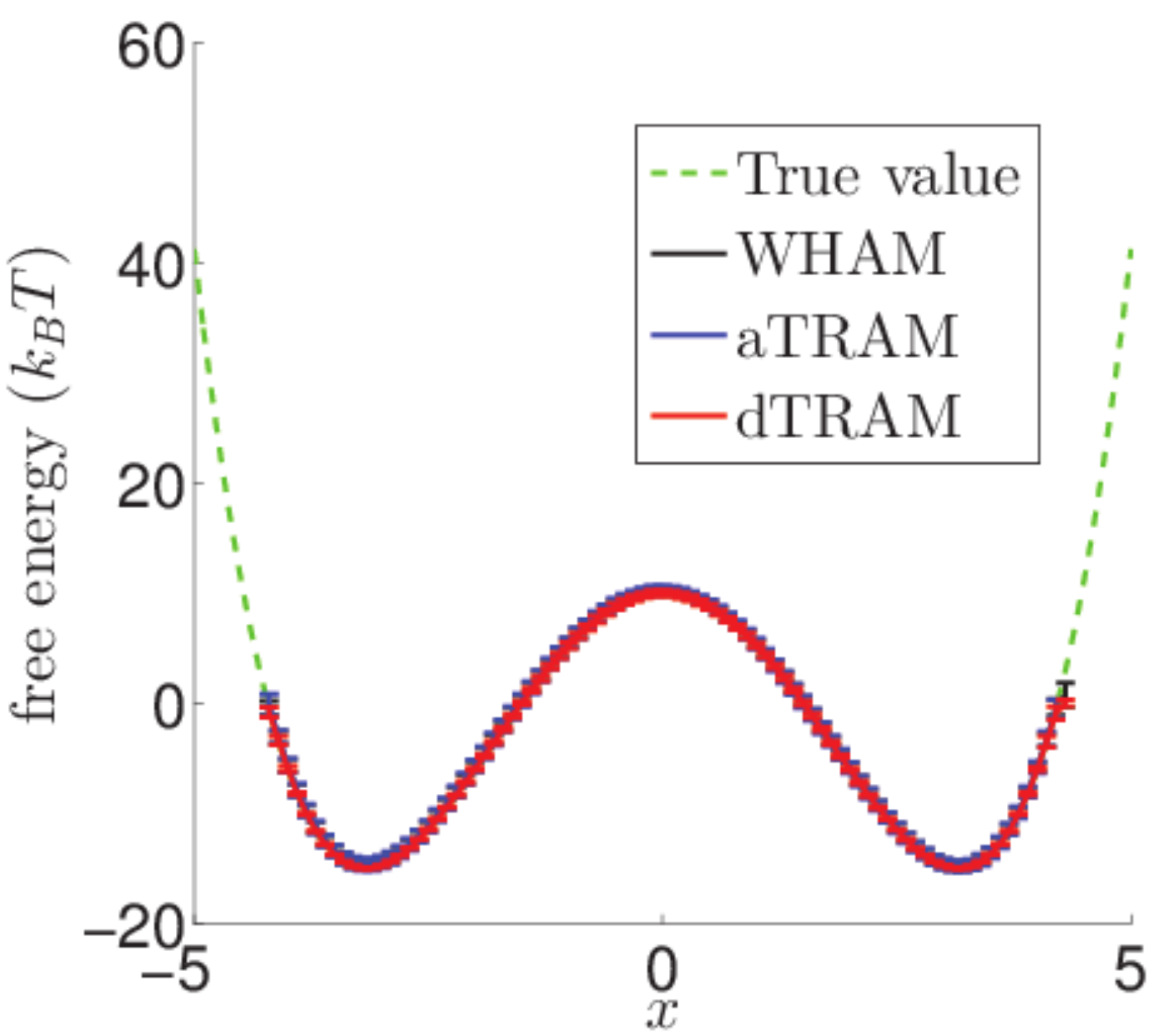}
\par\end{centering}

\protect\caption{\label{fig:US-doublewell}Estimation results of WHAM, the present
method (dTRAM), and our early approximate TRAM method (aTRAM) from
\cite{WuNoe_MMS14_TRAM1} based on umbrella sampling simulations in
a double-well potential. (A) Mean and standard deviation of the energy
barrier estimation error calculated over $30$ independent umbrella
sampling runs with $K=11$ umbrellas each. The $x$-Axis shows the
number of steps in an umbrella trajectory. (B) Mean and standard deviation
of estimates of the potential $u(x)$ generated by WHAM, dTRAM and
approximate TRAM using a trajectory length of $500$. (C) Same as
(B), but with a trajectory length of $10000$.}
\end{figure}

Here, we employ umbrella sampling simulations on a one-dimensional
double-well potential with a dimensionless energy:
\begin{equation}
u\left(x\right)=\frac{1}{4}x^{4}-5x^{2}-9.9874,
\end{equation}
illustrated in Fig.~\ref{fig:US-doublewell}B,C. The configuration
space is discretized using a one dimensional binning: All $x$-values
are assigned to the closest point from the set $\left\{ -5,-4.9,\ldots,4.9,5\right\} $,
generating up to 101 discrete states. However, in practice only about
80 states are visited and we exclude empty bins from the analysis.
Umbrella sampling simulations are conducted using $K=11$ different
biasing potentials given by:
\begin{equation}
b^{(k)}\left(x\right)=4\left(x-\bar{x}^{(k)}\right)^{2},\quad\text{for }k=1,\ldots,11
\end{equation}
where $\bar{x}^{(k)}=k-6$ is the center of the $k$-th biasing potential.
The dynamics are simulated using the Metropolis process described
in Appendix C. The bias potentials are chosen relatively weak, such
that the umbrella simulations near the transition state contain rare
events . In this case, WHAM is a poor estimator because it takes
relatively long to generate statistically independent samples.

In order to apply dTRAM, we evaluate all bias potentials for each
discrete state, compute the reweighting factors according to Eq. (\ref{eq:reweighting-factor-US}),
and store this information in a reweighting matrix. We additionally
store all state-to-state transition counts $c_{ij}^{(k)}$ for each
umbrella simulation $k$. Given this data the dTRAM estimation is
computed by iterating equations (\ref{eq:TRAM_iteration_v}-\ref{eq:TRAM_iteration_pi})
to convergence.

The performances of WHAM and dTRAM are compared in terms of the mean
error of the estimated energy barriers:
\begin{equation}
\mathrm{error}=\frac{1}{2}\left(\left|\Delta u_{AB}-\Delta u_{AB}^{\mathrm{approx}}\right|+\left|\Delta u_{BA}-\Delta u_{BA}^{\mathrm{approx}}\right|\right)
\end{equation}
where $\Delta u_{AB}$ and $\Delta u_{BA}$ are the energy barriers
for the $A\rightarrow B$ and $B\rightarrow A$ process, respectively,
and the superscript ``$\mathrm{approx}$'' represents the approximate
value obtained from the estimated $u(x)$. Fig.~\ref{fig:US-doublewell}A
compares the energy barrier estimation error using WHAM (black) and
dTRAM (red) as a function of the length of the umbrella simulations.
In addition, estimation results our earlier approximate TRAM method
from \cite{WuNoe_MMS14_TRAM1} are shown in blue. Fig.~\ref{fig:US-doublewell}B
and C show the energy profile estimated from the data using trajectory
lengths of $910$ and $10000$ steps per umbrella, respectively. All
means and standard deviations are obtained by repeating the simulation
30 times.

It is apparent that both estimators converge to the correct energy
barriers and energy profile in the limit of a large amount of simulation
data. For short simulations, the dTRAM estimates are significantly
better than the WHAM and approximate TRAM estimates - or in other
words much less simulation data is needed with dTRAM to obtain estimates
of equal quality than when WHAM or approximate TRAM are used. 

\begin{figure}[H]
\noindent \begin{centering}
\includegraphics[width=1\columnwidth]{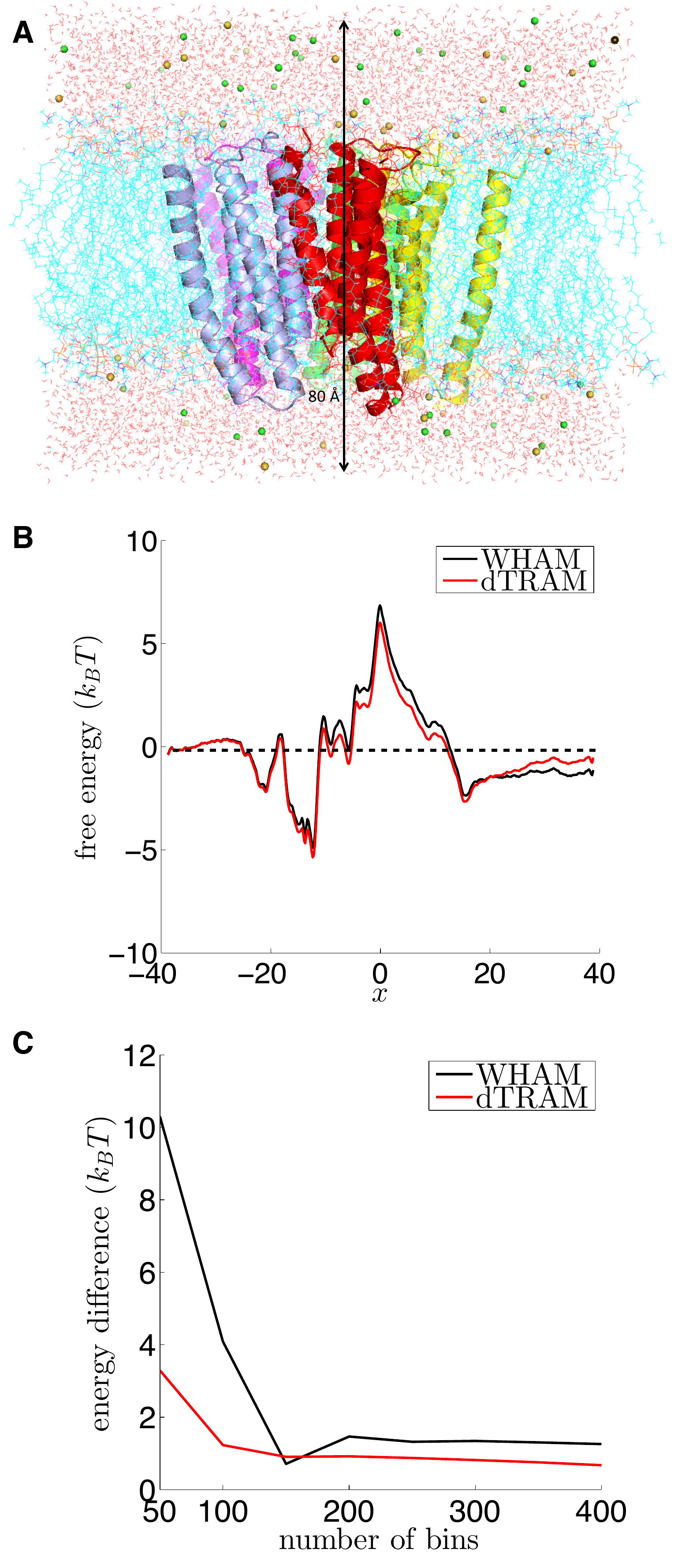} 
\par\end{centering}

\protect\caption{\label{fig:US-ion-channel}Estimation results of WHAM and dTRAM based
on umbrella sampling simulations of Na$^{+}$ passage through a GLIC
channel using simulations of Zhu and Hummer \cite{ZhuHummer_JCC12_ConvergenceWHAM}.
(A) Structure of the ion channel. (B) Free energy profile computed
by WHAM and dTRAM when using 400 bins to discretize the reaction coordinate
(membrane normal). (C) Systematic estimating error of the energies
of the end-states, measured by its difference from 0, as a function
of the number of discretization bins used.}
\end{figure}

In order to demonstrate the validity of dTRAM on molecular dynamics
data of a large protein system, we have used it to analyze umbrella
sampling simulations of the passage of Na$^{+}$ ions through the
transmembrane pore of the GLIC channel (Fig. \ref{fig:US-ion-channel}).
The data was generated in the simulations of Zhu and Hummer \cite{ZhuHummer_JCC12_ConvergenceWHAM}.
WHAM and dTRAM provide a similar free energy profile (Fig. \ref{fig:US-ion-channel}B),
but for a given number of bins used to discretize the membrane normal
used as a reaction coordinate, dTRAM provides a smaller systematic
error. The systematic error is measured in terms of the energy difference
calculated for the two end-states which should be 0 as a result of
the periodic boundary conditions used in the simulation (Fig. \ref{fig:US-ion-channel}C).

\section{Conclusions}

We have derived a maximum likelihood estimator for the transition-based
reweighting analysis method (TRAM). This estimator optimally combines
simulation trajectories produced at different thermodynamic states.
This is done by taking into account both the time correlations in
the trajectories via transition counts, and by considering that the
weight of every configuration can be given in any thermodynamic state
via the Boltzmann distribution. The present estimator operates a configuration
space discretization, and in particular also discretizes the reweighting
factors between different thermodynamic states. Hence we call the
present method discrete TRAM, or in short dTRAM.

dTRAM combines ideas from the weighted histogram analysis method (WHAM)
and reversible Markov state models (MSMs). We have shown that dTRAM
is in fact a proper generalization of both methods, i.e. both the
WHAM estimator and the reversible MSM estimator can be derived as
special cases from the dTRAM equations. Consequently, dTRAM can be
applied to any kind of simulation data that either WHAM or MSMs can
be applied to. In particular, dTRAM is useful for getting improved
estimates from general-ensemble simulations (replica-exchange molecular
dynamics, parallel or simulated tempering), multiple biased simulations
(umbrella sampling, metadynamics, conformational flooding etc). Like
MSMs, dTRAM is useful for obtaining estimates from swarms of short
simulations that are not in global equilibrium, but beyond MSMs it
can do so for trajectories produced under different thermodynamic
conditions, such as temperatures or Hamiltonians.

dTRAM provides estimates for both the equilibrium distribution at
a thermodynamic state of interest, and the kinetics at the thermodynamic
states simulated at. The kinetic estimates should be treated with
care as they are only useful when the data admits the choice of a
lagtime that is sufficiently long to parametrize a Markov model that
can predict long-term kinetics. For such estimates, it should be checked
whether they are converged as a function of the lag time, as it is
common practice in Markov state modeling. We will investigate the
suitability of dTRAM to estimate kinetic data in future work. In the
present paper we have only worked with short trajectory segments,
and hence have only used dTRAM in order to estimate equilibrium distributions
and free energies. It has been demonstrated that dTRAM provides superior
estimates of these properties compared to WHAM when applied to biased
simulations. 

The present reweighting formulation is straightforwardly applied when
the configuration space can be discretized in such a way that the
bias factor $\gamma_{i}^{(k)}$ is approximately constant within each
bin. Although such a discretization is easily done for umbrella sampling
while biasing one or two coordinates, it is unsuitable in other cases,
such as replica-exchange simulations. It has been suggested to construct
a joint discretization in configuration and energy space \cite{GallicchioLevy_JPCB05_TWHAM}.
An optimal solution to the problem will involve a self-consistent
evaluation of $\gamma_{i}^{(k)}$ in a way that integrates the sampled
configurations assigned to bin $i$ with a MBAR-like approach \cite{ShirtsChodera_JCP08_MBAR}. 

A python implementation of the dTRAM method is available under 
https://github.com/markovmodel/pytram.

\textbf{Acknowledgements}

The authors thank following colleagues for enlightening discussions:
John D. Chodera (MSKCC, New York), Gerhard Hummer (MPI for Biophysics,
Frankfurt), Alessandro Laio (SISSA, Trieste), Roland R. Netz (FU Berlin)
and the entire CMB team at FU Berlin. We thank Fangqiang Zhu for providing
us with the umbrella sampling simulation data of the GLIC ion channel.
This work was funded from following grants: WU 744/ 1-1 and DFG SFB
1114 (Wu), DFG SFB 740 and 958 (Mey), ERC grant ``pcCell'' (No\'{e}).

\selectlanguage{english}%
%

\selectlanguage{american}%
\clearpage

\section*{Appendix A: Discrete TRAM: Derivation and proofs}

\subsection*{A.1 Lagrange duality\label{sec:Lagrange-duality}}

A detailed description of Lagrange duality theory we refer to textbooks
\cite{ConvexAnalysisOptimization_bookI,ConvexAnalysisOptimization_bookII}.
Here we just summarize a few aspects of the theory relevant for solving
the TRAM problem.

Consider a constrained minimization problem of the following form:
\begin{equation}
\begin{array}{cl}
\min_{\mathbf{x}} & f\left(\mathbf{x}\right)\\
s.t. & f_{e}\left(\mathbf{x}\right)=0
\end{array}\label{eq:convex-problem}
\end{equation}
The Lagrangian function can be defined by
\begin{equation}
\Lambda\left(\mathbf{x},\mathbf{v}\right)=f\left(\mathbf{x}\right)+\mathbf{v}^{\intercal}f_{e}\left(\mathbf{x}\right)
\end{equation}
If (\ref{eq:convex-problem}) is a convex problem and some technical
assumption such as Slater condition holds, it can be proven that (\ref{eq:convex-problem})
is \emph{equivalent} to the following maximization problem: 
\begin{equation}
\max_{\mathbf{v}}g(\mathbf{v})\label{eq:dual-problem}
\end{equation}
where $g(\mathbf{v})$ is the Lagrange dual function defined by 
\begin{equation}
g(\mathbf{v}):=\min_{\mathbf{x}}\Lambda\left(\mathbf{x},\mathbf{v}\right)
\end{equation}
The equivalence here means the optimal values of the two problems
are equal, i.e., the solution $\mathbf{x}^{*}$ to (\ref{eq:convex-problem})
and the solution $\mathbf{v}^{*}$ to (\ref{eq:dual-problem}) satisfies
$g\left(\mathbf{v}^{*}\right)=f\left(\mathbf{x}^{*}\right)$. Moreover,
\begin{equation}
\mathbf{x}^{*}=\arg\min_{\mathbf{x}}\Lambda\left(\mathbf{x},\mathbf{v}^{*}\right)
\end{equation}

\subsection*{A.2 Single thermodynamic state}

We first consider the dual Lagrange approach to solving dTRAM at a
single thermodynamic state, i.e. the situation that is equivalent
with a reversible Markov state model. Here we first ignore the normalization
of the stationary distribution $\boldsymbol{\pi}$, i.e., $\sum_{i}\pi_{i}$
may not be equal to $1$. The maximum likelihood estimation of the
transition matrix $\mathbf{P}$ with a fixed equilibrium distribution
$\boldsymbol{\pi}$ is given by the following optimization problem:
\begin{equation}
\begin{array}{cl}
\min_{\mathbf{P}} & -L\left(\mathbf{P}\mid\mathbf{C}\right)\\
s.t. & \pi_{i}p_{ij}=\pi_{j}p_{ji}\\
 & \sum_{j}p_{ij}=1
\end{array}\label{eq:single-1}
\end{equation}
where
\begin{equation}
L\left(\mathbf{P}\mid\mathbf{C}\right)=\sum_{i,j}c_{ij}\ln p_{ij}
\end{equation}
is the log-likelihood function of $\mathbf{P}$. By using the Lagrange
duality theory, we have the following lemma:

\textbf{Lemma}: The minimization problem (\ref{eq:single-1}) is equivalent
to the maximization problem 
\begin{equation}
\begin{array}{crl}
\min_{\mathbf{v}} & h_{\mathbf{C}}\left(\boldsymbol{\pi},\mathbf{v}\right)= & \sum_{i,j}c_{ij}\ln\left(\pi_{i}v_{j}+\pi_{j}v_{i}\right)\\
 &  & -\sum_{i,j}c_{ij}\ln\pi_{j}-\sum_{i}v_{i}\\
 &  & -\sum_{i,j}c_{ij}\ln\left(c_{ij}+c_{ji}\right)+\sum_{i,j}c_{ij}
\end{array}\label{eq:single-2}
\end{equation}
and the optimal solution $\mathbf{P}^{*}$ to (\ref{eq:single-1})
and the optimal solution $\mathbf{v}^{*}$ to (\ref{eq:single-2})
satisfy
\begin{equation}
p_{ij}^{*}=\frac{\left(c_{ij}+c_{ji}\right)\pi_{j}}{\pi_{i}v_{j}^{*}+\pi_{j}v_{i}^{*}}
\end{equation}
\textbf{Proof}: The Lagrangian function of (\ref{eq:single-1}) is
defined as\textbf{
\begin{eqnarray}
\Lambda_{\boldsymbol{\pi}}\left(\mathbf{P},\boldsymbol{\lambda},\mathbf{v}\right) & = & -\sum_{i,j}c_{ij}\ln p_{ij}+\sum_{i,j}\lambda_{ij}\left(\pi_{i}p_{ij}-\pi_{j}p_{ji}\right)\nonumber \\
 &  & +\sum_{i}v_{i}\left(\sum_{j}p_{ij}-1\right)\nonumber \\
 & = & -\sum_{i,j}c_{ij}\ln p_{ij}+\sum_{i,j}\left(\pi_{i}\left(\lambda_{ij}-\lambda_{ji}\right)+v_{i}\right)p_{ij}\nonumber \\
 &  & -\sum_{i}v_{i}.\label{eq:Lagrangian-function}
\end{eqnarray}
}Note that
\begin{equation}
\frac{\partial\Lambda_{\boldsymbol{\pi}}}{\partial p_{ij}}=-\frac{c_{ij}}{p_{ij}}+\pi_{i}\left(\lambda_{ij}-\lambda_{ji}\right)+v_{i}
\end{equation}
Then the dual function of (\ref{eq:single-1}) is
\begin{eqnarray}
g_{\boldsymbol{\pi}}\left(\lambda,\nu\right) & = & \min_{\mathbf{P}}\Lambda_{\boldsymbol{\pi}}\left(\mathbf{P},\boldsymbol{\lambda},\mathbf{v}\right)\nonumber \\
 & = & \Lambda\left(\left[\frac{c_{ij}}{\pi_{i}\left(\lambda_{ij}-\lambda_{ji}\right)+v_{i}}\right],\boldsymbol{\lambda},\mathbf{v}\right)\nonumber \\
 & = & -\sum_{i,j}c_{ij}\ln\left(\frac{c_{ij}}{\pi_{i}\left(\lambda_{ij}-\lambda_{ji}\right)+v_{i}}\right)\nonumber \\
 &  & +\sum_{i,j}c_{ij}-\sum_{i}v_{i}\nonumber \\
 & = & \sum_{i,j}c_{ij}\ln\left(\pi_{i}\left(\lambda_{ij}-\lambda_{ji}\right)+v_{i}\right)-\sum_{i}v_{i}\nonumber \\
 &  & -\sum_{i,j}c_{ij}\ln c_{ij}+\sum_{i,j}c_{ij}\label{eq:dual-function}
\end{eqnarray}
and (\ref{eq:single-1}) is equivalent to the maximization problem
\begin{equation}
\max_{\boldsymbol{\lambda},\mathbf{v}}g_{\boldsymbol{\pi}}\left(\boldsymbol{\lambda},\mathbf{v}\right)
\end{equation}
From (\ref{eq:dual-function}), we can get
\begin{equation}
\frac{\partial g_{\boldsymbol{\pi}}\left(\boldsymbol{\lambda},\mathbf{v}\right)}{\partial\lambda_{ij}}=\frac{c_{ij}\pi_{i}}{\pi_{i}\left(\lambda_{ij}-\lambda_{ji}\right)+v_{i}}-\frac{c_{ji}\pi_{j}}{\pi_{j}\left(\lambda_{ji}-\lambda_{ij}\right)+v_{j}}
\end{equation}
and
\begin{equation}
\frac{\partial g_{\boldsymbol{\pi}}\left(\boldsymbol{\lambda},\mathbf{v}\right)}{\partial\lambda_{ij}}=0\Leftrightarrow\lambda_{ij}-\lambda_{ji}=\frac{c_{ij}\pi_{i}v_{j}-c_{ji}\pi_{j}v_{i}}{\left(c_{ij}+c_{ji}\right)\pi_{i}\pi_{j}}
\end{equation}
Therefore,
\begin{eqnarray}
\max_{\boldsymbol{\lambda}}g_{\boldsymbol{\pi}}\left(\boldsymbol{\lambda},\mathbf{v}\right) & = & \sum_{i,j}c_{ij}\ln\left(\pi_{i}\frac{c_{ij}\pi_{i}v_{j}-c_{ji}\pi_{j}v_{i}}{\left(c_{ij}+c_{ji}\right)\pi_{i}\pi_{j}}+v_{i}\right)\nonumber \\
 &  & -\sum_{i}v_{i}-\sum_{i,j}c_{ij}\ln c_{ij}+\sum_{i,j}c_{ij}\nonumber \\
 & = & \sum_{i,j}c_{ij}\ln\left(\frac{c_{ij}\left(\pi_{i}v_{j}+\pi_{j}v_{i}\right)}{\left(c_{ij}+c_{ji}\right)\pi_{j}}\right)\nonumber \\
 &  & -\sum_{i}v_{i}-\sum_{i,j}c_{ij}\ln c_{ij}+\sum_{i,j}c_{ij}\nonumber \\
 & = & \sum_{i,j}c_{ij}\ln\left(\pi_{i}v_{j}+\pi_{j}v_{i}\right)-\sum_{i,j}c_{ij}\ln\pi_{j}\nonumber \\
 &  & -\sum_{i}v_{i}-\sum_{i,j}c_{ij}\ln\left(c_{ij}+c_{ji}\right)+\sum_{i,j}c_{ij}\nonumber \\
 & = & h_{\mathbf{C}}\left(\boldsymbol{\pi},\mathbf{v}\right)
\end{eqnarray}
and the optimal value of (\ref{eq:single-1}) is equal to
\begin{equation}
\max_{\boldsymbol{\lambda},\mathbf{v}}g_{\boldsymbol{\pi}}\left(\boldsymbol{\lambda},\mathbf{v}\right)=\max_{\mathbf{v}}\left(\max_{\boldsymbol{\lambda}}g_{\boldsymbol{\pi}}\left(\boldsymbol{\lambda},\mathbf{v}\right)\right)=\max_{\mathbf{v}}h_{\mathbf{C}}\left(\boldsymbol{\pi},\mathbf{v}\right)
\end{equation}
and
\begin{eqnarray}
p_{ij}^{*} & = & \frac{c_{ij}}{\pi_{i}\left(\lambda_{ij}^{*}-\lambda_{ji}^{*}\right)+v_{i}^{*}}\nonumber \\
 & = & \frac{c_{ij}}{\pi_{i}\frac{c_{ij}\pi_{i}v_{j}^{*}-c_{ji}\pi_{j}v_{i}^{*}}{\left(c_{ij}+c_{ji}\right)\pi_{i}\pi_{j}}+v_{i}^{*}}\nonumber \\
 & = & \frac{\left(c_{ij}+c_{ji}\right)\pi_{j}}{\pi_{i}v_{j}^{*}+\pi_{j}v_{i}^{*}}
\end{eqnarray}
where
\begin{equation}
\lambda_{ij}^{*}-\lambda_{ji}^{*}=\frac{c_{ij}\pi_{i}v_{j}^{*}-c_{ji}\pi_{j}v_{i}^{*}}{\left(c_{ij}+c_{ji}\right)\pi_{i}\pi_{j}}
\end{equation}
We now consider the case that the stationary distribution is unknown.
In this case, the maximum likelihood estimate of $\pi$ is
\begin{eqnarray}
\boldsymbol{\pi}^{*} & = & \arg\min_{\boldsymbol{\pi}}\left(\min_{\mathbf{P}\text{ is feasible to \eqref{eq:single-1}}}-L\left(\mathbf{P}\right)\right)\nonumber \\
 & = & \arg\min_{\boldsymbol{\pi}}\left(\max_{\mathbf{v}}h_{\mathbf{C}}\left(\boldsymbol{\pi},\mathbf{v}\right)\right)
\end{eqnarray}

\subsection*{A.3 Multiple Simulations}

If we further consider multiple biased simulations with $\pi_{i}^{(k)}\propto\gamma_{i}^{(k)}\pi_{i}$,
we can also conclude that the maximum likelihood estimate of $\boldsymbol{\pi}$
can be given by the following min-max problem:
\begin{equation}
\min_{\boldsymbol{\pi}}\max_{\mathbf{v}^{(1)},\ldots,\mathbf{v}^{(K)}}\sum_{k}h_{\mathbf{C}^{(k)}}\left(\mathrm{diag}\left(\boldsymbol{\gamma}^{(k)}\right)\boldsymbol{\pi},\mathbf{v}^{(k)}\right)\label{eq:multiple-minimax-problem}
\end{equation}
For the sake of convenience, here we define $u_{i}=-\ln\pi_{i}$.
Then (\ref{eq:multiple-minimax-problem}) can be rewritten as
\begin{equation}
\min_{\mathbf{u}}\max_{\mathbf{v}^{(1)},\ldots,\mathbf{v}^{(K)}}\sum_{k}h_{\mathbf{C}^{(k)}}\left(\mathrm{diag}\left(\boldsymbol{\gamma}^{(k)}\right)\exp\left(-\mathbf{u}\right),\mathbf{v}^{(k)}\right)\label{eq:multiple-minimax-problem-u}
\end{equation}
Because $h\left(\exp\left(-\mathbf{u}\right),\mathbf{v}\right)$ is
a concave function of $\mathbf{v}$ if $\mathbf{u}$ is fixed and
a convex function of $\mathbf{u}$ if $\mathbf{v}$ is fixed, the
solution to (\ref{eq:multiple-minimax-problem-u}) is a saddle point
and characterized by (see Section 10.3.4 in \cite{BoydVandenberghe_Book_ConvexOptimization})
\begin{eqnarray}
\frac{\partial h_{\mathbf{C}^{(k)}}\left(\mathrm{diag}\left(\boldsymbol{\gamma}^{(k)}\right)\exp\left(-\mathbf{u}\right),\mathbf{v}^{(k)}\right)}{\partial\nu_{i}^{(k)}} & = & 0,\nonumber \\
\quad\quad\quad\quad i=1,\ldots,n,\: k=1,\ldots,K\\
\frac{\partial\sum_{k}h_{\mathbf{C}^{(k)}}\left(\mathrm{diag}\left(\boldsymbol{\gamma}^{(k)}\right)\exp\left(-\mathbf{u}\right),\mathbf{v}^{(k)}\right)}{\partial u_{i}} & = & 0,\nonumber \\
\quad\quad\quad\quad i=1,\ldots,n
\end{eqnarray}

Considering that
\begin{equation}
\frac{\partial h_{\mathbf{C}}\left(\mathrm{diag}\left(\boldsymbol{\gamma}\right)\exp\left(-\mathbf{u}\right),\mathbf{v}\right)}{\partial v_{i}}=\sum_{j}\frac{\left(c_{ij}+c_{ji}\right)\gamma_{j}\pi_{j}}{\gamma_{i}\pi_{i}v_{j}+\gamma_{j}\pi_{j}v_{i}}-1
\end{equation}
and
\begin{eqnarray}
 &  & \frac{\partial h_{\mathbf{C}}\left(\mathrm{diag}\left(\boldsymbol{\gamma}\right)\exp\left(-\mathbf{u}\right),\mathbf{v}\right)}{\partial u_{i}}\nonumber \\
 & = & \frac{\partial h_{\mathbf{C}}\left(\mathrm{diag}\left(\boldsymbol{\gamma}\right)\boldsymbol{\pi},\mathbf{v}\right)}{\partial\pi_{i}}\frac{\partial\pi_{i}}{\partial u_{i}}\\
 & = & -\pi_{i}\left(\sum_{j}\frac{\left(c_{ij}+c_{ji}\right)\gamma_{i}v_{j}}{\gamma_{i}\pi_{i}v_{j}+\gamma_{j}\pi_{j}v_{i}}-\frac{\sum_{j}c_{ji}}{\pi_{i}}\right)\nonumber \\
 & = & \sum_{j}c_{ji}-\sum_{j}\frac{\left(c_{ij}+c_{ji}\right)\gamma_{i}\pi_{i}v_{j}}{\gamma_{i}\pi_{i}v_{j}+\gamma_{j}\pi_{j}v_{i}}
\end{eqnarray}
We can conclude that the optimal solution to (\ref{eq:multiple-minimax-problem})
should satisfy
\begin{eqnarray}
\sum_{j}\frac{\left(c_{ij}^{(k)}+c_{ji}^{(k)}\right)\gamma_{j}^{(k)}\pi_{j}}{\gamma_{i}^{(k)}\pi_{i}v_{j}^{(k)}+\gamma_{j}^{(k)}\pi_{j}v_{i}^{(k)}} & = & 1\label{eq:constraint-nu}\\
\pi_{i}\sum_{j,k}\frac{\left(c_{ij}^{(k)}+c_{ji}^{(k)}\right)\gamma_{i}^{(k)}v_{j}^{(k)}}{\gamma_{i}^{(k)}\pi_{i}v_{j}^{(k)}+\gamma_{j}^{(k)}\pi_{j}v_{i}^{(k)}} & = & \sum_{j,k}c_{ji}^{(k)}\label{eq:constraint-pi}
\end{eqnarray}
This leads to the following iterative algorithm for unbiased estimation
of multiple simulations:
\begin{eqnarray}
v_{i}^{(k),\mathrm{new}} & = & v_{i}^{(k)}\sum_{j}\frac{\left(c_{ij}^{(k)}+c_{ji}^{(k)}\right)\gamma_{j}^{(k)}\pi_{j}}{\left(\gamma_{i}^{(k)}\pi_{i}v_{j}^{(k)}+\gamma_{j}^{(k)}\pi_{j}v_{i}^{(k)}\right)},\nonumber \\
 &  & \quad\quad\quad\quad\quad\quad\quad\quad\quad\quad\quad\text{for }i=1,\ldots n\label{eq:iterative-nu}\\
\pi_{i}^{\mathrm{new}} & = & \frac{\sum_{k,j}c_{ji}^{(k)}}{\sum_{k,j}\frac{\left(c_{ij}^{(k)}+c_{ji}^{(k)}\right)\gamma_{i}^{(k)}v_{j}^{(k)}}{\left(\gamma_{i}^{(k)}\pi_{i}v_{j}^{(k)}+\gamma_{j}^{(k)}\pi_{j}v_{i}^{(k)}\right)}},\nonumber \\
 &  & \quad\quad\quad\quad\quad\quad\quad\quad\quad\quad\quad\text{for }i=1,\ldots n\label{eq:iterative-pi}
\end{eqnarray}

\subsection*{A.4 Asymptotic correctness of dTRAM}

Here we show that dTRAM converges to the correct equilibrium distribution
and transition probabilities in the limit of large statistics. In
this limit, either achieved through long simulation trajectories or
many short simulation trajectories, the count matrices $\mathbf{C}^{(k)}=[c_{ij}^{(k)}]$
become:
\begin{equation}
c_{ij}^{(k)}=N_{i}^{(k)}\bar{p}_{ij}^{(k)}\label{eq:asymptotic-limit-counts}
\end{equation}
where $\bar{\mathbf{P}}^{(k)}=[\bar{p}_{ij}^{(k)}]$ is the matrix
of exact transition probabilities (no statistical error), which satisfies
\begin{equation}
\bar{\pi}_{i}\gamma_{i}^{(k)}\bar{p}_{ij}^{(k)}=\bar{\pi}_{j}\gamma_{j}^{(k)}\bar{p}_{ji}^{(k)}
\end{equation}
where $\boldsymbol{\pi}=[\bar{\pi}_{i}]$ are the exact stationary
probabilities of configuration states. Inserting (\ref{eq:asymptotic-limit-counts})
into the dTRAM equations (\ref{eq:dTRAM_equation1}-\ref{eq:dTRAM_equation2}),
we obtain:
\begin{eqnarray}
 &  & \sum_{k}\sum_{j}\frac{\left(c_{ij}^{(k)}+c_{ji}^{(k)}\right)\gamma_{i}^{(k)}\bar{\pi}_{i}N_{j}^{(k)}}{\gamma_{i}^{(k)}\bar{\pi}_{i}N_{j}^{(k)}+\gamma_{j}^{(k)}\bar{\pi}_{j}N_{i}^{(k)}}\\
 & = & \sum_{k}\sum_{j}\frac{\left(N_{i}^{(k)}\bar{p}_{ij}^{(k)}+N_{j}^{(k)}\bar{p}_{ji}^{(k)}\right)\gamma_{i}^{(k)}\bar{\pi}_{i}N_{j}^{(k)}}{\gamma_{i}^{(k)}\bar{\pi}_{i}N_{j}^{(k)}+\gamma_{j}^{(k)}\bar{\pi}_{j}N_{i}^{(k)}}\\
 & = & \sum_{k}\sum_{j}\frac{\left(N_{i}^{(k)}\gamma_{j}^{(k)}\bar{\pi}_{j}+N_{j}^{(k)}\gamma_{i}^{(k)}\bar{\pi}_{i}\right)\bar{p}_{ji}^{(k)}N_{j}^{(k)}}{\gamma_{i}^{(k)}\bar{\pi}_{i}N_{j}^{(k)}+\gamma_{j}^{(k)}\bar{\pi}_{j}N_{i}^{(k)}}\\
 & = & \sum_{k}\sum_{j}\bar{p}_{ji}^{(k)}N_{j}^{(k)}\\
 & = & \sum_{k}\sum_{j}c_{ji}^{(k)}
\end{eqnarray}
and thus the first dTRAM equation is satisfied. Furthermore, we obtain:
\begin{eqnarray}
 &  & \sum_{j}\frac{\left(c_{ij}^{(k)}+c_{ji}^{(k)}\right)\gamma_{j}^{(k)}\bar{\pi_{j}}}{\gamma_{i}^{(k)}\bar{\pi}_{i}N_{j}^{(k)}+\gamma_{j}^{(k)}\bar{\pi}_{j}N_{i}^{(k)}}\\
 & = & \sum_{j}\frac{\left(N_{i}^{(k)}\bar{p}_{ij}^{(k)}+N_{j}^{(k)}\bar{p}_{ji}^{(k)}\right)\gamma_{j}^{(k)}\bar{\pi_{j}}}{\gamma_{i}^{(k)}\bar{\pi}_{i}N_{j}^{(k)}+\gamma_{j}^{(k)}\bar{\pi}_{j}N_{i}^{(k)}}\\
 & = & \sum_{j}\frac{N_{i}^{(k)}\bar{p}_{ij}^{(k)}\gamma_{j}^{(k)}\bar{\pi_{j}}+N_{j}^{(k)}\bar{p}_{ij}^{(k)}\gamma_{i}^{(k)}\bar{\pi_{i}}}{\gamma_{i}^{(k)}\bar{\pi}_{i}N_{j}^{(k)}+\gamma_{j}^{(k)}\bar{\pi}_{j}N_{i}^{(k)}}\\
 & = & \sum_{j}\bar{p}_{ij}^{(k)}\\
 & = & 1
\end{eqnarray}
and thus the second dTRAM equation is satisfied as well. From the
above two equations, we can conclude that in the statistical limit
(either achieved by long trajectories or many short trajectories),
the solution of the dTRAM equations converges to the correct equilibrium
distribution and the correct transition probabilities. Note that we
have assumed that all trajectory data is in local equilibrium within
each starting state $i$ - if this is not the case the counts $c_{ij}$
and thus also the estimated $\pi_{i}$ and $p_{ij}$ will be biased.
Thus, if the data is of such a nature that local equilibrium is a
concern (e.g. metadynamics or uncoupled short MD trajectories), all
estimates should be computed as a function of the lag time $\tau$.

\section*{Appendix B: Double-well simulation model}

The simulation trajectory $\{x_{t}^{(k)}\}_{t=0}^{M}$ is generated
by a Metropolis simulation model, which is a reversible Markov chain
with initial state $x_{0}^{(k)}=\bar{x}^{(k)}$, stationary distribution
$\pi_{i}^{(k)}\propto\mathrm{e}^{-u_{i}-b_{i}^{(k)}}$, and transition
probability
\begin{eqnarray}
 &  & \Pr\left(x_{t+1}^{(k)}=x'|x_{t}^{(k)}=x\right)\\
 &  & =\left\{ \begin{array}{ll}
\min\left\{ \frac{\exp\left(-u\left(x'\right)-v^{(k)}\left(x'\right)\right)}{\exp\left(-u\left(x\right)-v^{(k)}\left(x\right)\right)}q\left(x|x'\right),q\left(x'|x\right)\right\} , & x'\neq x\\
1-\sum_{y\neq x}\Pr\left(x_{t+1}^{(k)}=y|x_{t}^{(k)}=x\right), & x'=x
\end{array}\right.\nonumber 
\end{eqnarray}
where $q\left(x'|x\right)$ denotes the proposal distribution which
satisfies $q\left(x'|x\right)\propto1_{\left|x'-x\right|\le0.2}$
and $\sum_{x'}q\left(x'|x\right)=1$.

\end{document}